\documentclass[aps,prd,amssymb,eqsecnum]{revtex4}

\usepackage{bm}
\usepackage{graphicx}

\begin{document}
\draft

%\twocolumn[\hsize\textwidth\columnwidth\hsize\csname
%@twocolumnfalse\endcsname

\rightline{
\large\baselineskip20pt\rm\vbox to20pt{
\baselineskip14pt
\hbox{OCU-PHYS-221}
\hbox{AP-GR-20}}}

\title{Quasinormal Ringing for Acoustic Black Holes at Low Temperature}
\author{
Hiroyuki Nakano,
Yasunari Kurita,
Kouji Ogawa, 
Chul-Moon Yoo %,
}
\address{
Department of Mathematics and Physics, Graduate School of Science,
Osaka City University, Osaka 558-8585, Japan
}
\date{\today}

\begin{abstract}
We investigate a condensed matter ``black hole'' analog, 
taking the Gross-Pitaevskii (GP) equation as a starting point. 
The linearized GP equation corresponds to a wave equation on a black hole background, 
giving quasinormal modes under some appropriate conditions. 
We suggest that we can know the detailed characters 
and corresponding geometrical information about the acoustic black hole 
by observing quasinormal ringdown waves in the low temperature condensed matters. 
\end{abstract}

%\pacs{04.70.Bw, 47.90.+a}
%\vspace{2ex}

\maketitle
%]

%**********************************************************************
%**********************************************************************
%**********************************************************************

%%%%%%%%%%%%%%%%%%%%%%%%%%%%%%%%%%%%%%%%%%%%%%%%%%%%%%%%%%%%%%%%%%%%%
\section{Introduction}\label{sec:1}
%%%%%%%%%%%%%%%%%%%%%%%%%%%%%%%%%%%%%%%%%%%%%%%%%%%%%%%%%%%%%%%%%%%%%

Black holes are one of the most fascinating objects for theoretical physics, especially for General Relativity. 
Strikingly, they have thermodynamic properties based on the Bekenstein-Hawking entropy formula and 
the Hawking's renowned effect~\cite{hawking74}, which might be a touchstone of quantum theory of gravity. 
Unfortunately, however, it seems to be impossible to detect the Hawking radiation from 
realistic astrophysical black holes because the Hawking temperature of astrophysical black holes is so low that 
its intensity is too weak to be detected.
For the purpose of confirming the Hawking radiation, 
it will be greatly helpful to perform laboratory experiments using some kinds of analogy.

There are a lot of works about an acoustic ``black hole'' as the
experimental field for Hawking radiation.
The acoustic black hole has a sound horizon corresponding to an event horizon of a black hole spacetime. 
Unruh~\cite{unruh81} originally suggested a hydrodynamic analog of an event horizon~\cite{misner+73} 
as a more accessible phenomenon which might shed some light on the Hawking effect. 
In principle, an event horizon for sound waves appears wherever 
there is a closed surface through which a fluid flows inwards at the speed of sound. 
In the vicinity of the surface, the flow is subsonic on one side  
and supersonic on the other side; 
there is a close analogy between sound propagation on 
a background hydrodynamic flow and field propagation in a curved spacetime. 
Thus, if such an analogous black hole is materialized, 
it will be experimentally powerful tool for confirming the Hawking radiation. 
There are some works in this direction~\cite{Barcelo:2001ca,Sakagami:2001ph}.  
Recently, the subject has attracted much 
attention~\cite{visser98,Berti:2004ju,Cardoso:2004fi,Kim:2004sf,Lepe:2004kv,Cadoni:2004my}.

In order to detect the Hawking radiation in the experimental laboratory, thermal 
fluctuations will become main 
obstacles.
Therefore we consider an acoustic black hole in the Bose-Einstein condensation (BEC) system 
at low temperature.
Thanks to recent technological advances, the BEC 
has been observed in 1995 
in a remarkable series of experiments on vapors of rubidium~\cite{Anderson95} and sodium~\cite{Davis95} 
in which the atoms were confined in magnetic traps and cooled down to extremely low temperature, 
of the order of fractions of microkelvins. (See the review paper~\cite{Dalfovo} about the phenomenon of 
Bose-Einstein condensation of dilute gases from a theoretical perspective.) 
Garay {\it et al}.~\cite{Garay:1999sk,Garay:2000jj} have analyzed in detail the case of a condensate in a ring trap, 
and proposed a realistic scheme for adiabatically creating stable acoustic black/white holes. 
It seems to be possible to materialize an acoustic black hole in the BEC system.

Astrophysical black holes will obey the uniqueness theorem for black hole spacetimes,
and there will be no hair other than mass, angular momentum and electric charge. 
On the other hand, it will be possible to realize analogous black holes having many hairs.
Since analogous black holes have geometrical correspondence, it means that many kinds of geometry
will be realized in the analogous systems. 
Thus, it is necessary to identify which corresponding classical geometry is realized in the experiments.

Geometries of acoustic black holes are decided by physical situations  
(e.g. the equations of state, %potentials of fields and 
boundary conditions on fluid systems and so on ). 
Since what directly affects sound waves or Hawking radiations is its background geometry or wave equation
on the analogous curved spacetime, it is needed to confirm 
whether the required geometry
is materialized or not by detecting some kind of a physical phenomenon 
which is not determined by factors except for the geometry.

Without this confirmation, we can not clean away the fear of false geometries 
or false analogies.

One of the ways to do it is to detect damped oscillating waves of fluid fluctuations 
under some perturbation.
We make a point that the modes of damped oscillations in the linear phase 
are uninfluenced by given perturbation   % initial fluctuations, 
and directly depend only on the geometry.

This damped oscillating waves are well known as quasinormal ringdown waves for astrophysical 
black holes context.
The first order perturbed Einstein equation gives a wave equation, and 
quasinormal modes (QNMs) are complex frequency solutions of the wave equation with a purely 
outgoing-wave boundary condition at infinity and ingoing-wave at the horizon, with vanishing incoming-wave amplitude. 
They have geometrical information outside the horizon and, in fact, the quasinormal ringdown waves 
have astrophysical significance and are searched by using the data 
from the gravitational wave detectors~\cite{Tsunesada:2004}.
In this paper, we analyze quasinormal ringdown waves of an acoustic black hole 
with the procedures which are frequently used in analysis about 
those of astrophysical ones, 
and using the QNMs of this wave, we discuss the distinguishability of the geometry.

On the other hand,
there is a subtle problem concerning the detection of QNMs in gravitational astrophysics.
For the purpose of designing the gravitational wave detector or making the data analysis, 
we must predict theoretically amplitudes of gravitational waves which we want to detect. 
The amplitude depends on the energy which is extracted by the gravitational waves from sources 
such as supernovae or neutron star mergers. 
Thus it is important to know
how many energies of the process transport to the gravitational waves.
When it comes to quasinormal ringdown waves, 
however, it depends on the final phase of the gravitational wave emission process such as black hole formation; 
hence we need to clarify all the emission processes including nonlinear phases 
(e.g. the merger phase of binary neutron star systems) for the  estimation of the amplitudes. 
Generally, this theoretical estimation can be done only with complicated 
numerical works which is not yet sufficiently accurate at the present time. 
Some data analysis has been done on the assumption that 
the energy transport is a few percent to the background black hole mass, 
however this assumption does not have much foundation \cite{Tsunesada:2004}. 
If analogous  black holes are realized in some experimental laboratories, 
we will obtain some suggestion about the energy extraction rate by QNMs.
Therefore the study of QNMs in acoustic black holes is also significant
 in this direction. 

In this paper, we consider an analogous black hole in a Bose-Einstein condensate 
and predict the quasinormal ringdown waves in the system for the purpose of 
investigating their corresponding classical geometry. 
The BEC system has some remarkable properties that 
it is a quantum mechanical system and its dynamical evolution obeys the Gross-Pitaevskii (GP) equation, 
in which any profile of its density can be materialized freely by controlling the exterior potential in actual experiments. 
The linearized GP equation gives a wave equation which corresponds to that on a black hole background, 
giving QNMs under some appropriate conditions. 
In order to derive the quasinormal frequencies, we will use the WKB method given by Schutz and Will \cite{SW}. 
Some quasinormal frequencies in the case where the section area of a nozzle is constant and 
the background density has some combinations of `power-form' will be shown. 
This choice of the background density covers for all scale invariant forms of the density, where 
the power of the density is a unique parameter for the system. 
This fact will be helpful for actual experiments.

The paper is organized as follows. In Sec.~\ref{sec:GPE}, 
we summarize the Gross-Pitaevskii equation which is the equation of motion 
of a Bose-Einstein condensed matter, and the linearized Gross-Pitaevskii equation. 
And then, an acoustic black hole which has 
an event horizon for sound wave is shown by considering the differential equation. 
The detailed analysis has been discussed 
by C.~Barcelo, S.~Liberati and M.~Visser~\cite{Barcelo:2000tg} 
and we summarize it in the Appendix. 
For this black hole, we calculate the QNMs by using 
the WKB approach in Sec.~\ref{sec:QNM}. 
Finally, we show how to confirm the geometrical information 
and discuss the remaining problem in Sec.~\ref{sec:Dis}.

%%%%%%%%%%%%%%%%%%%%%%%%%%%%%%%%%%%%%%%%%%%%%%%%%%%%%%%%%%%%%%%%%%%%%
\section{Gross-Pitaevskii Equation and Acoustic Black Hole}\label{sec:GPE}
%%%%%%%%%%%%%%%%%%%%%%%%%%%%%%%%%%%%%%%%%%%%%%%%%%%%%%%%%%%%%%%%%%%%%

Now, in order to consider a Bose-Einstein condensate, 
we introduce the field operator $\psi$ which is divided into two parts as 
\begin{eqnarray}
\psi = \Psi + \tilde{\psi} \,,
\nonumber 
\end{eqnarray}
where $\Psi$ is called as the order parameter 
and $\tilde{\psi}$ is some perturbation around the order parameter. 
The order parameter is a complex function 
defined as the expectation value of the field operator, 
and written as the following form: 
\begin{eqnarray}
\Psi = \sqrt{n_0} \,e^{i\,\varphi}.
\label{eq:Psi}
\end{eqnarray}
Here, the functions $n_0$ and $\varphi$ denote a condensate density 
and a phase factor of the condensed matter, respectively. 
The order parameter is a classical field called the 
``wave function of the condensate''. 
In the following, we ignore $\tilde{\psi}$ and consider only this order parameter. 

When the number of atoms is large and the atomic interactions are 
sufficiently small in the vicinity of zero temperature, 
the Gross-Pitaevskii equation is the equation of motion 
of a condensed matter
\begin{eqnarray}
i \hbar \frac{\partial \Psi}{\partial t} 
= -\left(\frac{\hbar^2}{2m}\nabla^2+V_{\rm ext}\right) \Psi 
+ \frac{4\,\pi\,a\,\hbar^2}{m} |\Psi|^2 \Psi \,,
\end{eqnarray}
where $m$ is the mass of the atoms, $a$ is the scattering length and  
the atoms are trapped by an external potential $V_{\rm ext}({\mathbf{x}})$
\cite{Dalfovo}. This has the form of a ``nonlinear Schr\"odinger equation''.
Substituting Eq.~(\ref{eq:Psi}) into the above equation, 
we obtain two equations from the real and imaginary parts, respectively: 
\begin{eqnarray}
\frac{\partial n_0}{\partial t}
&=& -\nabla (n_0 \,\nabla \Phi)  \,, 
\label{eq:cont} \\ 
\frac{\partial \Phi}{\partial t} 
&=& -\frac{1}{2}(\nabla \Phi)^2 
+ \frac{\hbar^2}{4m^2} 
\left(\frac{\nabla^2 n_0}{n_0} 
- \frac{(\nabla n_0)^2}{2 n_0^2} \right) 
- \frac{V_{\rm ext}}{m} - \frac{4\,\pi\,a\,\hbar^2}{m^2} n_0 \,. 
\label{eq:Ber}
\end{eqnarray}
Here we can introduce a new function 
$\Phi = \frac{\hbar}{m} \varphi \,,$ 
and the velocity of a condensed matter ${\bf v}_0$ as 
\begin{eqnarray}
{\bf v}_0 &=& \nabla \Phi\,,
\end{eqnarray}
and above two equations (\ref{eq:cont}) and (\ref{eq:Ber}) 
seem to be the continuity and the Bernoulli equations, respectively. 

Next, we derive the equation 
for the linear perturbation of the Gross-Pitaevskii equations (\ref{eq:cont}) 
and (\ref{eq:Ber}) 
\begin{eqnarray}
\tilde n_0 &=& n_0(1+\zeta) \,, \nonumber \\ 
\tilde \Phi &=& \Phi + \eta \,.
\end{eqnarray}
Here we consider $n_0$ and $\Phi$ as a background, and 
$\zeta$ and $\eta$ represent small perturbations around them. 
For simplicity, we assume that the background is stationary and independent of time. 
Then, the background values $n_0$ and $\Phi$ are 
obtained by solving the following time-independent Gross-Pitaevskii equation:  
\begin{eqnarray}
\nabla (n_0 \,\nabla \Phi) &=& 0 \,, 
\label{eq:contB} \\ 
-\frac{1}{2}(\nabla \Phi)^2 
+ \frac{\hbar^2}{4m^2} 
\left(\frac{\nabla^2 n_0}{n_0} 
- \frac{(\nabla n_0)^2}{2 n_0^2} \right) 
- \frac{V_{\rm ext}}{m} - \frac{4\,\pi\,a\,\hbar^2}{m^2} n_0 &=& 0 \,.
\label{eq:BerB}
\end{eqnarray}
On the other hand, the linear perturbation equations are given by 
\begin{eqnarray}
\frac{\partial \zeta}{\partial t} +{\bf v}_0 \cdot \nabla \zeta 
+\nabla^2 \eta
+\nabla \ln n_0 \cdot \nabla \eta &=& 0 \,, 
\label{eq:Pcont} \\
\frac{\partial \eta}{\partial t} +{\bf v}_0 \cdot \nabla \eta 
-\frac{\hbar^2}{4 m^2} \frac{1}{n_0} 
\nabla (n_0 \nabla \zeta) 
+ \frac{4\,\pi\,a\,\hbar^2}{m^2} n_0 \zeta &=& 0 \,.
\label{eq:PBer} 
\end{eqnarray}
Now we assume that the variation of the density $n_0$ 
can be neglected for the healing length ($=1/\sqrt{8\pi a n_0}$), i.e., 
the third term on the L.H.S. of Eq.~(\ref{eq:PBer}) is neglected. 
This approximation is considered as a low frequency approximation. 
And then, we obtain the simple equation: 
\begin{eqnarray}
\frac{\partial \eta}{\partial t} +{\bf v}_0 \cdot \nabla \eta 
+ \frac{4\,\pi\,a\,\hbar^2}{m^2} n_0 \zeta &=& 0 \,.
\label{eq:PBerc} 
\end{eqnarray}
Eliminating the density perturbation $\zeta$ 
from Eqs.~(\ref{eq:Pcont}) and (\ref{eq:PBerc}), we obtain 
a second order differential equation for the phase (velocity) perturbation $\eta$  
\begin{eqnarray}
\left(\frac{\partial}{\partial t} +{\bf v}_0 \cdot \nabla \right)
\left[ \frac{1}{c_s^2}
\left(\frac{\partial}{\partial t} +{\bf v}_0 \cdot \nabla \right)
\eta \right] 
- \frac{1}{n_0} \nabla (n_0 \nabla \eta) = 0 \,,
\label{eq:VPE}
\end{eqnarray}
where we have defined $c_s$ as the sound velocity: 
\begin{eqnarray}
c_s = \frac{\hbar}{m} \sqrt{4\,\pi\,a\,n_0} \,. 
\end{eqnarray}
It should be noted that the sound velocity $c_s$ is a local quantity and
also that if this low frequency approximation is not taken into account, 
Eq.(\ref{eq:Pcont}) and (\ref{eq:PBer}) can not be reduced to a single second order differential equation 
for the condensate phase perturbation. 
We will give some discussion about this approximation in Sec.~\ref{sec:Dis}. 

In the following, for simplicity, we consider the case 
where the background velocity has only $x$-component, 
i.e., $\{v_0^i\}=\{v_0,\,0,\,0\}$ where the superscript $i$ denotes $\{x,\,y,\,z\}$. 
From Eq.~(\ref{eq:VPE}), we can read off the distance 
between two nearby points in the corresponding curved spacetime as 
\begin{eqnarray}
ds^2 &=& \frac{\alpha n_0}{c_s} 
\left(-(c_s^2 - v_0^2)dt^2 -2 v_0 dt dx
+dx^2 + dy^2 + dz^2 \right) \nonumber \\ 
&=& \frac{\alpha n_0}{c_s} 
\biggl(-(c_s^2 - v_0^2)\left(dt + \frac{v_0}{c_s^2 - v_0^2} dx \right)^2
+ \frac{c_s^2}{c_s^2 - v_0^2} dx^2  + dy^2 + dz^2 \biggr) \nonumber \\ 
&=& \frac{\alpha n_0}{c_s} 
\left(-\left(1 - \frac{v_0^2}{c_s^2}\right) c_s^2\,d\tau^2 
+ \displaystyle{\frac{1}{1 - \displaystyle{\frac{v_0^2}{c_s^2}}}} dx^2 
+ dy^2 + dz^2 \right) \,. 
\label{eq:metriK}
\end{eqnarray}
This detailed derivation is shown in the Appendix. 
Here, we have introduced the new time coordinate $\tau$ defined by 
\begin{eqnarray}
d\tau = dt + \frac{v_0}{c_s^2 - v_0^2} dx \,.
\end{eqnarray}
When the superfluid velocity changes from subsonic 
to supersonic velocity, there is a sound horizon for the direction of the $x$-axis. 
We shall call this system with the horizon an acoustic black hole, 
because this has an event horizon for sound waves.

%%%%%%%%%%%%%%%%%%%%%%%%%%%%%%%%%%%%%%%%%%%%%%%%%%%%%%%%%%%%%%%%%%%%%
\section{Quasinormal modes}\label{sec:QNM}
%%%%%%%%%%%%%%%%%%%%%%%%%%%%%%%%%%%%%%%%%%%%%%%%%%%%%%%%%%%%%%%%%%%%%

In this section, we consider the background configuration of a condensed matter 
which has a sound horizon and then derive the quasinormal frequencies in this background. 
In order to obtain a sound horizon, 
we consider the one-dimensional background flow of a condensed matter 
with the velocity $v_0$ in the $x$-direction. 
The background density $n_0$ is derived by using the continuity equation (\ref{eq:contB}). 
Then, we can calculate the external potential by Eq.~(\ref{eq:BerB}). 
In practice, if the external potential is chosen as the above value, 
it is possible to construct an acoustic black hole. 
We note that the background discussed in the following 
has a constant section area of a nozzle. This is a single black hole case. 

Using Eq.~(\ref{eq:PET}), the linear perturbation equation of the system is written as 
\begin{eqnarray}
\left[ 
- \partial_{t}^2 - 2 \,v_0 \partial_t \,\partial_x 
- \frac{c_s^2}{n_0} \left( \partial_x \frac{n_0\, v_0}{c_s^2} \right) \partial_t 
+ \frac{c_s^2}{n_0} \partial_x \left(
n_0 \left(1-\frac{v_0^2}{c_s^2}\right) \partial_x \right) 
\right] \eta = 0 \,.
\end{eqnarray}
For the above equation, using the background density $n_0$, 
we rewrite the sound velocity $c_s$ as 
\begin{eqnarray}
c_s^2 = C^2\,n_0 \,, 
\end{eqnarray}
and the background velocity $v_0$ can be expressed as
\begin{eqnarray}
v_0 = \frac{V}{n_0} \,.
\label{eq:den-rel}
\end{eqnarray}
Here Eq.~(\ref{eq:den-rel}) is obtained from the continuity equation (\ref{eq:cont}) 
and the assumption that the section area of a nozzle is constant. 
Then we rewrite the above perturbation equation into the following form: 
\begin{eqnarray}
\left[ 
- \partial_{t}^2 - \frac{2\,V}{n_0} \partial_t \,\partial_x 
- V \left( \partial_x \frac{1}{n_0} \right) \partial_t 
+ C^2 \partial_x \left(
n_0 \left(1-\frac{V^2}{C^2}\frac{1}{n_0^3}\right) \partial_x \right) 
\right] \eta = 0 \,.
\label{eq:pet2}
\end{eqnarray}
We assume that the background is static, 
so it is convenient to consider the Fourier coefficients of
the phase (velocity) perturbation $\eta$
\begin{eqnarray}
X_{\omega}(x)  = \int dt \,\eta(t,\,x) \,e^{i\,\omega\,t} \,. 
\end{eqnarray}
In order to simplify the perturbation equation, i.e., 
to remove the first differentiation with respect to $t$, 
we introduce the following auxiliary functions, $F_{\omega}$ and $G_{\omega}$: 
\begin{eqnarray}
X_{\omega}(x) &=& e^{i\,F_{\omega}(x)}\,G_{\omega}(x) \,,
\end{eqnarray}
where the function $F_{\omega}$ has been chosen as 
\begin{eqnarray}
F_{\omega}(x) &=& - \int^x dx\, \frac{\omega\,V}{(C^2\,n_0^2 - V^2/n_0)} \,.
\end{eqnarray}
Then, the perturbation equation takes the following form:
\begin{eqnarray}
\left[ 
\frac{\left (n_0^3 - \beta^2 \right)^2}{n_0^5} \partial_x^2 
+ \frac{\left( \partial_x n_0\right) \left( n_0^3 + 2\,\beta^2\right) 
\left (n_0^3 - \beta^2 \right)}{n_0^6} \partial_x
+ \sigma^2
\right] G_{\omega} = 0 \,,
\end{eqnarray}
where we have defined $\beta=V/C$ and $\sigma=\omega/C$, respectively. 

Next, in order to obtain the Schr\"odinger-type equation,
we introduce the new coordinate $x_*$ as 
\begin{eqnarray}
\frac{dx_*}{dx} &=& \frac{1}{f(x)},
\nonumber \\ 
f(x) &=& \frac{n_0^3 - \beta^2}{n_0^{5/2}} \,.
\end{eqnarray}
From the above coordinate transformation, the perturbation equation becomes 
\begin{eqnarray}
\left[ \partial_{x_*}^2 
+ \frac{1}{2}\frac{\left( \partial_x n_0\right) \left (n_0^3 - \beta^2 \right)}
{n_0^{7/2}} \partial_{x_*}
+ \sigma^2
\right] G_{\omega} = 0 \,.
\end{eqnarray}
Furthermore, in order to remove the first differentiation with respect to $x_*$, 
we replace the function $G_{\omega}$ as
\begin{eqnarray}
G_{\omega}(x) &=& n_0^{-1/4}\,H_{\omega}(x) \,.
\end{eqnarray}
Then we obtain the Schr\"odinger-type equation for the function $H_{\omega}$ as
\begin{eqnarray}
\left[ \partial_{x_*}^2 + \left(\sigma^2 - {\cal V}(x_*)\right)
\right] H_{\omega} &=& 0 \,,
\nonumber \\ 
{\cal V}(x_*) &=& \frac{1}{4}\frac{\left(n_0^3 - \beta^2 \right)^2
\left( \partial_x^2 n_0\right)}{n_0^6} 
-\frac{1}{16}\frac{\left(n_0^3 - \beta^2 \right)
\left(n_0^3 - 13\,\beta^2 \right)\left( \partial_x n_0\right)^2}{n_0^7} \,. 
\label{eq:final-pet}
\end{eqnarray}
In order to calculate the quasinormal frequencies, 
it is necessary to choose the appropriate potential ${\cal V}$. 
For example, we consider the case that the background density $n_0$ has the following form:
\begin{eqnarray}
n_0 = N x^a \,,
\label{eq:nbeki}
\end{eqnarray}
where $a$ is a constant. From Eq.~(\ref{eq:den-rel}), the background velocity becomes 
\begin{eqnarray}
v_0 = \frac{V}{N}x^{-a} \,.
\end{eqnarray}
It is noted that the different choice of $a$ derives 
the different relative velocity profile in the fluid flow. 
In this case, the horizon arises at 
\begin{eqnarray}
x = \left( \frac{\beta^{2/3}}{N} \right)^{1/a} \,,
\end{eqnarray}
because of the sound velocity $c_s = C \sqrt{N} x^{a/2}$. 

\begin{figure}[ht]
\includegraphics[scale=.3]{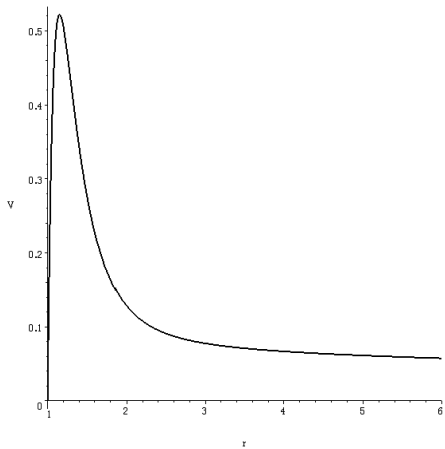}
\caption{The function ${\cal V}(x)$ in the case of $n_0=x^{5/3}$. 
${\cal V}(x)=0$ at $x=1$ (horizon) and the limit $x \rightarrow \infty$. }
\label{fig:ptl}
\end{figure}

We find that the range of parameter $a$ should be $-2/5 < a <2$ for 
the condition in which the potential ${\cal V}$ vanishes at infinity. 
Furthermore, in order to obtain the potential 
which goes to zero at the horizon of an acoustic black hole, 
it is necessary to restrict $a$ to $0<a<2$. 
This condition for $a$ also ensures the short-range character of the potential ${\cal V}$ 
which has been mentioned in \cite{ChandrasekharDetweiler1975}:
\begin{eqnarray}
\int^{\infty}_{-\infty} {\cal V}(x_*) dx_* < \infty \,.
\end{eqnarray}
Thus, in the following, we consider the case of the parameter range $0<a<2$.
To obtain the above background density $n_0$, 
we may choose the external potential $V_{\rm ext}$ of the
background Gross-Pitaevskii equation as one calculated by using Eq.~(\ref{eq:BerB}). 
In the following discussion, we choose $\beta=1$ and $N=1$. 
For example, in the case of $a=5/3$, 
substituting $n_0$ given by Eq.~(\ref{eq:nbeki}) into the perturbation equation 
(\ref{eq:final-pet}), 
we obtain the simple potential ${\cal V}$ and the coordinate transformation $f$ as 
\begin{eqnarray}
{\cal V} &=& \frac{5}{48} \frac{(x^5-1)(x^5+19)}{x^{31/3}},
\nonumber \\ 
f &=& \frac{x^5-1}{x^{25/6}} \,.
\end{eqnarray}
The potential ${\cal V}$ shown in Fig.~\ref{fig:ptl} 
has one peak and goes to zero at $x_* \rightarrow \pm \infty$ in the $x_*$ coordinate. 

In order to obtain the quasinormal frequencies of the background with the above potential, 
we use the usual WKB approach. (See Schutz and Will~\cite{SW} 
in the case of a black hole. The higher order WKB approach 
has been developed in Refs.~\cite{Iyer:1986np,Iyer:1986nq,Konoplya:2003ii}.) 
The quasinormal frequencies are obtained as 
\begin{eqnarray}
\sigma^2 = {\cal V}_0
- i \left(n+\frac{1}{2}\right)(-2\,{\cal V}_0^{''})^{1/2} \,, 
\end{eqnarray}
where $n$ is the non-negative integer for $\Re(\sigma)>0$, $n$ is the negative integer for $\Re(\sigma)<0$, 
and the prime (${}^{'}$) denotes the differentiation with respect to $x_*$. 
Here $n$ indices the $n$-th QNM, and the subscript $0$ on the potential denotes the value at 
\begin{eqnarray}
\,x_0 = 1.152 \quad {\mbox{where}} \quad
{\cal V}_0^{'}=0 \,.
\end{eqnarray}
The differentiation of the potential in the above calculation are obtained directly. 
The quasinormal frequencies are shown in Table~I. 

\begin{figure}[ht]
\includegraphics[scale=.3]{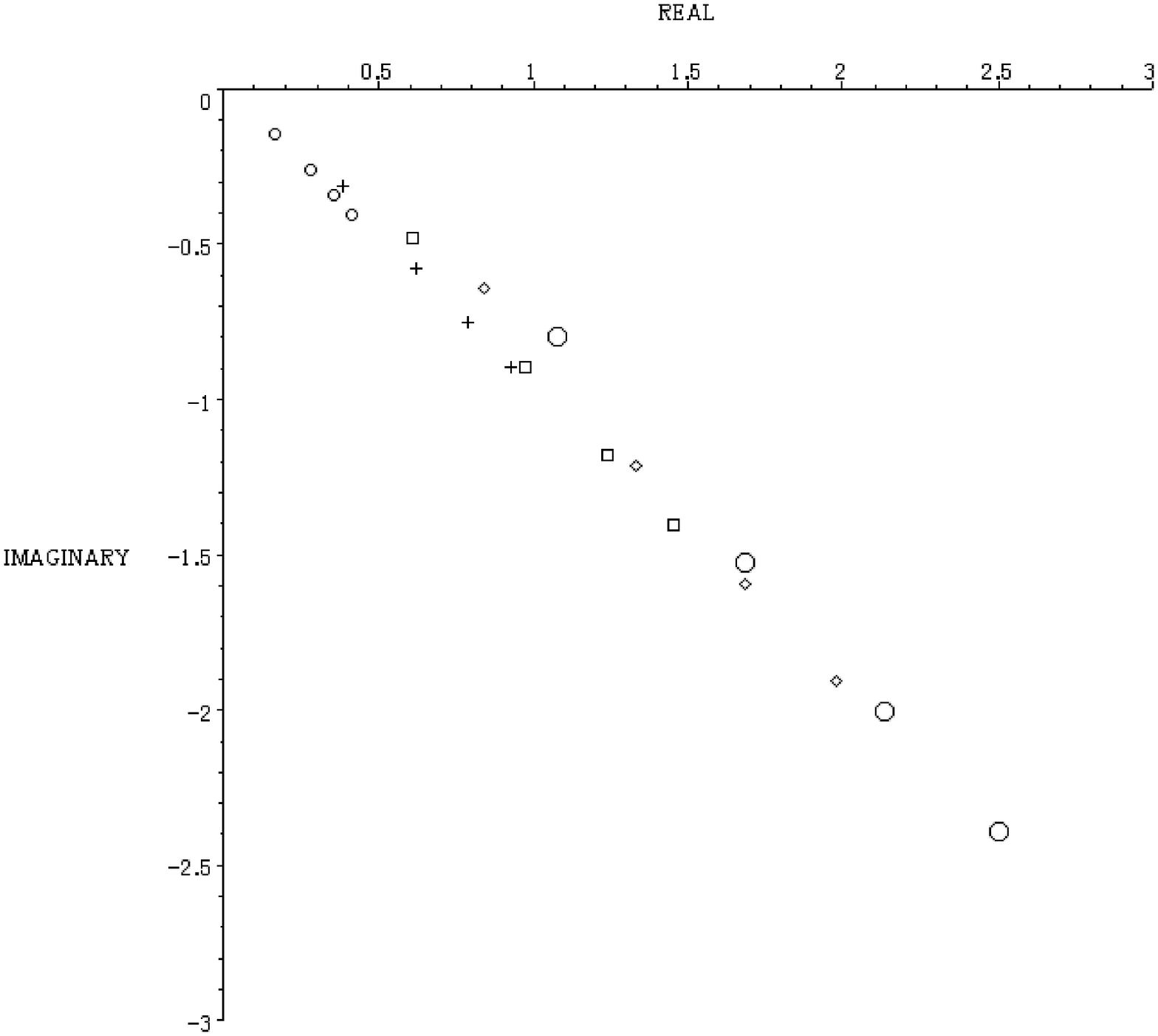}
\caption{The quasinormal frequencies in the case of $n_0=x^{a}$ for various $a$. 
The symbols $\circ$, {\tiny $+$}, {\tiny $\Box$}, {\tiny $\Diamond$} 
and $\bigcirc$ show the case of $a=1/3,\,2/3,\,1,\,4/3$ 
and $5/3$, respectively.}
\label{fig:bekidep}
\end{figure}

Now, we have a interest in the power $a$ dependence of the quasinormal frequencies.
The quasinormal frequencies in the cases of $a=1/3,\,2/3,\,1,\,4/3$ and $5/3$
are summarized in Fig.~\ref{fig:bekidep}. 
The figure shows the least damped mode in each value of $a$ differs each other, and thus, 
we can distinguish the power $a$ by the quasinormal frequencies and check the realization of 
the corresponding geometry in the analogous black hole system. 

\begin{figure}[ht]
\includegraphics[scale=.3]{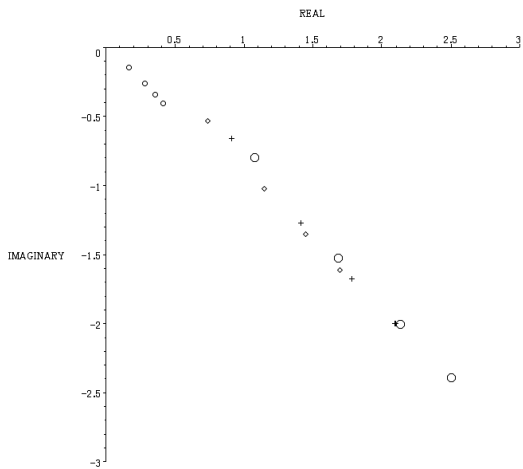}
\caption{The quasinormal frequencies in the case of $n_0=(b\,x+(1-b)x^5)^{1/3}$. 
The symbols $\bigcirc$, {\tiny $+$}, {\tiny $\Diamond$} 
and $\circ$ show the case of $b=0,\,1/3,\,2/3$ and $1$, respectively.}
\label{fig:2beki}
\end{figure}

In Fig.~\ref{fig:2beki}, the QNMs in the case of $n_0=(b\,x+(1-b)x^5)^{1/3}$ are shown, 
where the parameter $b$ takes the value of $0,\,1/3,\,2/3$ and $1$. 
In this case, the sonic horizon exists at the same position ($x=1$) as in the previous case of $n_0=x^a$. 
We find that as $b \to 1$, the QNMs change into those of $n_0=x^{1/3}$ continuously. 

\begin{table}[t]
 \begin{center}
  \begin{tabular}{c|cc}
   $n$ & $\Re(\sigma)$ & $\Im(\sigma)$ \\ \hline
   $0$ & 1.075 & -0.796 \\
   $1$ & 1.685 & -1.523 \\
   $2$ & 2.132 & -2.006 \\
   $3$ & 2.501 & -2.394 \\
  \end{tabular}
\label{table:lowQ}
\caption{Some quasinormal frequencies for $a=5/3$.}
 \end{center}
\end{table}

%%%%%%%%%%%%%%%%%%%%%%%%%%%%%%%%%%%%%%%%%%%%%%%%%%%%%%%%%%%%%%%%%%%%%
\section{Discussion}\label{sec:Dis}
%%%%%%%%%%%%%%%%%%%%%%%%%%%%%%%%%%%%%%%%%%%%%%%%%%%%%%%%%%%%%%%%%%%%%

In this paper, we have discussed the QNMs of an acoustic black hole 
in the context of the Bose-Einstein condensate. 
This acoustic black hole is prepared by choosing the external potential, 
and a single black hole case has been discussed.
Using the WKB method \cite{SW}, 
we have derived some quasinormal frequencies in the case 
where the background density has a power form or some combination of power forms.

The QNMs in this system can be considered as the ringdown waves 
emitted from a perturbed acoustic black hole as in the usual astrophysics context.
We can obtain the detailed characters and corresponding geometrical 
information about the acoustic black hole from its detection as follows. 
It is our purpose mentioned in Sec. \ref{sec:1}. 
From the quasinormal frequencies shown in Fig.~\ref{fig:bekidep} and Fig.~\ref{fig:2beki}, 
we can identify the density of the 
BEC which determine the potential function (\ref{eq:final-pet}). 
Thus, they are useful to check whether the system is realized to our expectations in an actual experiment
and to know the potential function (\ref{eq:final-pet}) and 
the corresponding geometrical information such as Eq.~(\ref{eq:curvedmetric}). 
Not only the least damped mode but
the higher damped modes are useful to identify the corresponding classical geometry.

It is noted that $|\Re(\sigma)/\Im(\sigma)| \sim 1$. 
This means that the ringdown waves from this system have a short decaying time. 
In astrophysical black holes, the situation is similar: their ringdown waves have a short decaying time.
One exception occurs in the case of extremal black holes whose Hawking temperature equals zero.
Thus, at the present moment, we do not have a strong interest in such black holes.

We also note that ``Schwarzschild like black holes'' can not be obtained 
in the fluid configuration
discussed in this paper. 
However the Schwarzschild black hole is one of the most impotant objects in astrophysical 
black holes context. 
If the acoustic black hole which has the same geometry as the Schwarzschild black hole 
is realized in some experiment, it is quite meaningful. 
Why we could not obtain the geometry 
is because the sound velocity $c_s$ and the background velocity $v_0$ are related to the density $n_0$ 
as Eq.~(\ref{eq:den-rel}). 
In order to obtain a Schwarzschild black hole in the BEC, we have to consider other fluid configurations~\cite{purple}. 
And then, the quasinormal frequencies are solvable by using the Leaver's method~\cite{Leaver} in which 
we may solve the three-term recurrence relation.

In our study, we have considered the low frequency approximation for the linearized Gross-Pitaevskii equation. 
This is relevant in the region where the wavelength of sounds 
is much longer than the healing length of the system. 
The quasinormal ringdown waves obtained in this paper satisfy the condition and 
therefore, our investigation is appropriate. 

The QNMs obtained in this paper depend on the outer geometry from the sound horizon. 
The higher damped modes which are not obtained in this paper will depend on the geometry in the vicinity of 
the horizon, where, however, the wavelength becomes shorter and shorter due to 
the blue shift effect associated with the horizon.  
When we intend to obtain the higher damped modes and, correspondinly, the information about horizon, 
we cannot use the low frequency approximation 
and have to investigate the full linearized Gross-Pitaevskii equation.

In this case, we cannot obtain a single second order differential equation for the condensate 
phase perturbation,
and sound waves will break the causality 
because their propagation speed may become infinity. 
There are many papers about this ``broken'' Lorentz invariance problem~\cite{jacobson91,unruh95,Corley:1996ar,Corley98}. 

Recently, Konoplya studied the influence of the back reaction of the Hawking radiation upon QNMs
in the case of (2+1) dimensional black hole~\cite{Konoplya:2004ik}.
Such effects might be more important in this BEC system because the influence will give 
some information about the Hawking radiation at low temperature.

We leave these issues for future investigations.

%===========================%
\acknowledgements
%===========================%

We would like to thank H. Ishihara, K. Nakao and H. Kozaki for fruitful discussions. 
We also thank M. Tsubota for useful discussions.
HN is supported by the JSPS Research Fellowships
for Young Scientists, No.~5919.

%===========================%
%===========================%
%===========================%

\appendix

%\begin{eqnarray}
%\frac{\partial \sqrt{n_0}}{\partial t}
%&=& -\frac{\hbar}{2m}\left(2 \nabla \sqrt{n_0} \cdot \nabla \varphi 
%+\sqrt{n_0} \nabla^2 \varphi\right)  \,, \\ 
%\hbar \frac{\partial \varphi}{\partial t} 
%&=& -\frac{\hbar^2}{2m}\left((\nabla \varphi)^2 
%- \frac{\nabla^2 \sqrt{n_0}}{\sqrt{n_0}} \right) 
%- V_{\rm ext} 
%- \frac{4\,\pi\,a\,\hbar^2}{m} n_0 \,.
%\end{eqnarray}

%%%%%%%%%%%%%%%%%%%%%%%%%%%%%%%%%%%%%%%%%%%%%%%%%%%%%%%%%%%%%%%%%%%%%
\section{Analogue Gravity in the Bose-Einstein Condensation}\label{app:AG}
%%%%%%%%%%%%%%%%%%%%%%%%%%%%%%%%%%%%%%%%%%%%%%%%%%%%%%%%%%%%%%%%%%%%%

In this appendix, we summarize the derivation 
of the curved metric (\ref{eq:metriK}) from the wave equation (\ref{eq:VPE}). 
More detailed analysis has been discussed in \cite{Barcelo:2000tg}.

Using the background continuity equation (\ref{eq:contB}), 
we rewrite the wave equation (\ref{eq:VPE}) for the linear perturbation 
of a velocity potential as
\begin{eqnarray}
&& \frac{1}{n_0} 
\left(\frac{\partial}{\partial t} + (\nabla \cdot {\bf v}_0) \right) 
\frac{n_0}{c_s^2}
\left(\frac{\partial}{\partial t} +{\bf v}_0 \cdot \nabla \right)\eta 
- \frac{1}{n_0} \nabla (n_0 \nabla \eta) = 0 \,.
\label{eq:PET}
\end{eqnarray}
Now, let us compare the above equation with the scalar wave equation in a general curved spacetime: 
\begin{eqnarray}
\Box \eta = \frac{1}{\sqrt{-g}} \frac{\partial}{\partial x^{\mu}}
\left[\sqrt{-g} g^{\mu\nu} \frac{\partial}{\partial x^{\nu}} \eta 
\right] \,,
\label{eq:covariantDar}
\end{eqnarray}
where $g^{\mu\nu}$ denotes a metric and $g$ is its determinant. 
We choose the coordinate as $\{x^{\mu}\}=\{t,x,y,z\}$, 
and if we set the metric components as 
\begin{eqnarray}
\sqrt{-g} g^{\mu\nu} = 
\alpha \, \left(\begin{array}{cc}
-\displaystyle{\frac{n_0}{c_s^2}} & 
-\displaystyle{\frac{n_0\,v_0^i}{c_s^2}} \\
-\displaystyle{\frac{n_0\,v_0^j}{c_s^2}} & 
n_0 \left(\delta^{ij}-\displaystyle{\frac{v_0^i v_0^j}{c_0^2}} \right)
\end{array}\right) \,,
\end{eqnarray}
where the superscript $i$ denotes $\{x,\,y,\,z\}$,
then the wave equation (\ref{eq:covariantDar}) is equivalent to (\ref{eq:PET}).
Using the value of determinant $\sqrt{-g}=\alpha^2 n_0^2 /c_s$, 
the inverse matrix is calculated as 
\begin{eqnarray}
g_{\mu\nu} = 
\frac{\alpha n_0}{c_s} \, \left(\begin{array}{cc}
-(c_s^2 - {\bf v_0}\cdot{\bf v_0}) & -v_0^i \\
{} & {} \\ 
-v_0^j & \delta^{ij}
\end{array}\right) \,.
\label{eq:curvedmetric}
\end{eqnarray}
Thus we can interpret that the perturbed system characterized by Eq.(\ref{eq:PET}) corresponds to
the wave propagation on the curved spacetime background with the metric (\ref{eq:curvedmetric}).

%************************************************************************
%************************************************************************

%************************************************************************
%************************************************************************

\end{document}